\documentclass[12pt,preprint2]{aastex}       
\usepackage{emulateapj5}
\usepackage{onecolfloat5}

\newcommand{\ale}{\lesssim}
\newcommand{\age}{\gtrsim}

\newcommand{\rxj}{RX~J052600.3$-$660433}
\newcommand{\sgr}{SGR~0526$-$66}
\newcommand{\chandra}{{\it Chandra}}
\newcommand{\mc}{\multicolumn}
\newcommand{\expnt}[2]{\ensuremath{#1 \times 10^{#2}}}   
\newcommand{\rosat}{\textit{ROSAT}}
\newcommand{\hst}{\textit{HST}}

\shorttitle{The Quiescent Counterpart of \sgr}
\shortauthors{Kulkarni et al.}

\slugcomment{Accepted by ApJ}

\begin{document}

\twocolumn[

\title{The Quiescent Counterpart of the Soft Gamma-ray Repeater \sgr}
\author{
S.~R.~Kulkarni,
D.~L.~Kaplan
}
\affil{Department of Astronomy 105-24,
         California Institute of Technology, Pasadena, CA 91125 USA.}
\email{srk@astro.caltech.edu}
\author{H.~L.~Marshall}
\affil{Center for Space Research, Massachusetts Institute of Technology,
         Cambridge, MA 02139, USA.}
\author{D.~A.~Frail}
\affil{National Radio Astronomy Observatory, P. O. Box O,
        Socorro, NM 87801, USA.}
\author{
T.~Murakami \&\ D.~Yonetoku
}
\affil{Department of Physics, Kanazawa University, Kadoma-cho,
         Kanazawa, 920-1192, Japan}


\begin{abstract}
It is now commonly believed that Soft gamma-ray repeaters (SGRs) and
Anomalous X-ray pulsars (AXPs) are magnetars --- neutron stars powered
by their magnetic fields.  However, what differentiates these two
seemingly dissimilar objects is, at present, unknown.  We present
\chandra\ observations of \rxj, the quiescent X-ray counterpart of
\sgr, famous for the intense burst of 5 March 1979.  The source is
unresolved at the resolution of \chandra.  Restricting to a period
range around 8~s, the period noted in the afterglow of the burst of 5
March 1979, we find evidence for a similar periodicity in two epochs
of data obtained 20 months apart. The secular period derivative based
on these two observations is $6.6(5)\times 10^{-11}\,$s$^{-1}$,
similar to the period derivatives of the magnetars.  As is the case
with other magnetars, the spectrum is best fitted by a combination of
a black body and a power law.  However, quite surprisingly, the photon
index of the power law component is $\Gamma\sim 3$ --- intermediate to
those of AXPs and SGRs.  This continuum of $\Gamma$ leads us to suggest
that the underlying physical parameter which differentiates SGRs from
AXPs is manifested in the power law component. Two decades ago, \sgr\ was a
classical SGR whereas now it behaves like an AXP. Thus it is possible
that the same object cycles between SGR and AXP state.  We speculate
that the main difference between AXPs and SGRs is the geometry of the
$B$-fields and this geometry is time dependent.  Finally, given the
steep spectrum of \rxj, the total radiated energy of \rxj\ can be much
higher than traditionally estimated. If this energy is supplied by the
decay of the magnetic field then the inferred $B$-field of \rxj\ is in
excess of $10^{15}\,$G, the traditional value for
magnetars. Independent of this discussion, there could well be a class
of neutron stars, $10^{14}\ale B \ale 10^{15}\,$G, which are neither
radio pulsars nor magnetars.
\end{abstract}


]

\section{Introduction}
\label{sec:introduction}

The soft gamma-ray repeater \sgr\ played a key role in our
understanding of high energy transients. It was from this source that an
intense burst was observed on 5 March, 1979 \citep{mgia+79,cdpt+80}.
The burst was followed  by an ``afterglow'' emission with an apparent
8-s periodicity.  The source of the burst was quickly localized to the
supernova remnant N49 (also known as SNR 0525$-$66.1) in the Large
Magellanic Cloud \citep{eklc+80}.  Observations with \rosat\ identified
a quiescent and bright ($L_X\sim 10^{36}\,$ erg s$^{-1}$) X-ray
counterpart, \rxj\ \citep*{rkl94}.

The intense burst of 5 March 1979 and the luminous afterglow with 8-s
periodicity provided the first and strongest evidence for super-strong
magnetic field strengths, $B\sim 10^{15}\,$G. Such strong fields are
needed to both confine the radiating plasma as well as allow the
radiation to escape \citep{dt92,paczynski92}.
However, such highly magnetized neutron stars or ``magnetars'' were
originally motivated by theoretical considerations --- namely
strong convection would naturally lead to growth of magnetic
fields during the process of the collapse of the proto-neutron
star core \citep{dt92,td93}.

Separately, another group of neutron stars, the so-called Anomalous
X-ray pulsars (AXPs), were recognized as a new class of neutron stars
(\citealt*{pth95}; \citealt{ms95}). The AXPs were noted for a narrow period
distribution, between 6 and 20 s; luminous X-ray emission, $L_X\sim
10^{35}\,$erg and apparent lack of a donor star. The sources were
``anomalous'' in that the source of the quiescent emission was neither
rotational (from the known $\dot P$) nor accretion (apparent lack of
companion).  Various authors speculated and suggested
that AXPs are also magnetars --- specifically, their X-ray
emission to arise from the decay of a magnetar-like field strength
\citep{td93}.

\begin{deluxetable}{r  c c c c l l}
\tablecaption{Position of \rxj}
\tablewidth{0pt}
\tablehead{
\colhead{ObsId} & & \colhead{$x$} & \colhead{$y$} & &
\colhead{$\alpha-05^{\rm h}26^{\rm m}$} &
\colhead{$\delta+66\degr04\arcmin$} \\ \cline{3-4} \cline{6-7}
 & & \mc{2}{c}{(pixels)} & & \colhead{(sec)} & \colhead{(arcsec)} \\
}
\startdata
747  && 4160.596(8) & 4135.884(8) && 00.8791(6) & $-$36.180(4)\\
1957 && 4090.665(8) & 4025.371(8) && 00.9094(6) & $-$36.424(4)\\
2515 && 4091.27(3)  & 4025.06(3)  && 00.911(4) &  $-$36.45(1)\\
\tableline
Average && & && 00.8948(4) & $-$36.307(3) \\
\enddata
\tablecomments{Positions are J2000.  The values in parentheses above
  are 1-$\sigma$ statistical uncertainties.  There is an additional 1-$\sigma$
position uncertainty of $\approx 0\farcs6$ in each coordinate due to
aspect uncertainties.}
\label{tab:position}
\end{deluxetable}

The discovery of periodicity in SGRs \citep{kds+98} and the overlap
of $P$ and $\dot P$ between AXPs and SGRs continued to motivate
a unified magnetar framework for both these objects.
In particular, the magnetic field strength inferred from $P$
and $\dot P$ (vacuum dipole framework) led to estimates of
about $10^{14}\,$G for both these objects, within a factor of few
of that estimated for AXPs and SGRs.

Toward the end of nineties, thanks to large area radio
pulsar searchers, astronomers became aware of a growing group of radio pulsars
\citep{ckl+00} with similarly long periods and with inferred magnetic
field strengths approaching $10^{14}$\,G (hereafter HBPSRs).  These
pulsars possess no special attributes linking them to either the AXPs
(no steady bright quiescent X-ray emission; \citealt*{pkc00}) or the
SGRs (no bursting history).  Thus periodicity alone does not appear to
be a sufficient attribute for classification.

Nonetheless, the recent discovery of bursts of radiation
--- similar to the minor bursts seen from SGRs --- from two
AXPs are strong empirical confirmation of a link
between AXPs and SGRs (\citealt*{gkw02}; \citealt{kg02}).  However, we are
still at a loss what specific physical parameter[s] differentiates AXPs
from SGRs.

One plausible notion is that AXPs and SGRs are linked temporally.
Specifically, three out of the six AXPs are associated with supernova
remnants (SNRs) whereas only \sgr\ has a plausible SNR association
\citep{gsgv01}.  Taken at face value, these data suggest that AXPs
evolve into SGRs.   However, this hypothesis has two 
problems.  First, the rotational periods of SGRs are similar to those
of AXPs, about 10-s.  Second, inferred magnetic field strengths of
SGRs are similar to (and perhaps even larger than) those of AXPs
\citep{hurley99,mereghetti99}.  Thus, there is no strong period
or B-field evolution between the two groups.

In our opinion, the above two objections are sufficiently severe that
we must continue searching for underlying physical parameter[s] that
differentiates between AXPs and SGRs.  To this end, investigating the
properties of the quiescent emission, which in practice means
spectroscopic and rotational properties, appear promising.  
Here we report investigation of the quiescent X-ray emission of \sgr,
comparing and contrasting the quiescent emission with those of AXPs
and other SGRs.

\section{Observations and Analysis}
We observed \rxj\ thrice with the \textit{Chandra X-ray Observatory}
\citep{wtvso2000}.  Our goal was to search for periodicity and obtain
broad-band spectrum of \rxj. To this end the first two observations
were obtained with a high temporal resolution.  Specifically, the
back-side illuminated ACIS-S3 charge coupled device (CCD) was used in
a 1/8 sub-array mode with a frame read every 0.44104 s.  The first
observation (ObsId 747) began at 4.02 January 2000; the total
on-source integration time was 37.2\,ks.  The second observation (ObsId
1957) used the same CCD setup and started on 31.94 August 2001; the
total integration time was 46.5 ks.  The last observation (ObsId 2515)
was designed to image the entire SNR and hence we used the entire S3
chip with a frame time of 3.2\,s and an integration time of
6.8\,ks.

All data sets were processed identically.  First, we reprocessed the
level-1 event data with the \texttt{CIAO} tool
\texttt{acis\_process\_events} to account for updated gain maps and
geometric calibration of the
spacecraft\footnote{\url{http://asc.harvard.edu/ciao/threads/geom\_par/}}.
We then produced a level-2 event file by copying only events with the
correct
grades\footnote{\textit{ASCA} grades 0, 2, 3, 4, and 6.}.
With this file, we restricted the data to the energy range of
0.3--10~keV, and filtered out times of high background count-rates.
Finally, we barycentered the data with the \texttt{axBary} tool using
the position of the \rxj\ (\S \ref{sec:pos}).

\subsection{Image Analysis}
\label{sec:pos}
\begin{figure}[b]
\plotone{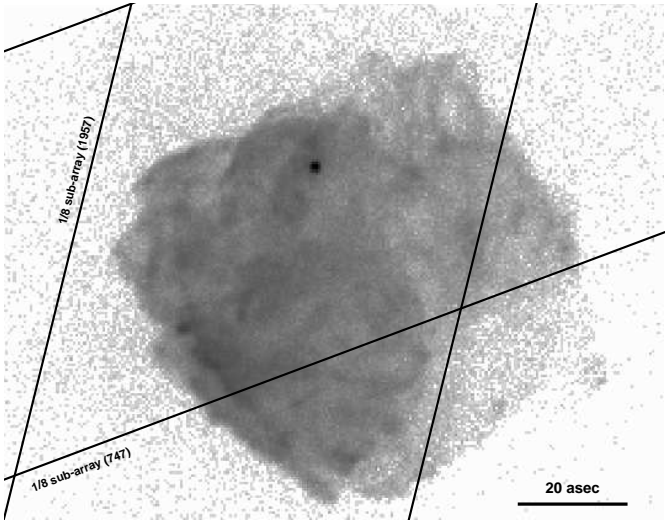}
\caption[] {Image of the supernova remnant (SNR) N49 obtained from the
\chandra\ X-ray satellite.  This image is a composite of all three
observations binned to 1-pixel ($0\farcs49$) resolution.  \rxj, the
quiescent counterpart to \sgr, is the point source towards the top.  The limited
spatial regions covered by the sub-arrays (ObsID's 747 and 1957) are
indicated by the parallel lines.  A $20\arcsec$ scale is shown, and
the orientation follows the usual convention with North up and East to
the left.
\label{fig:n49-image} }
\end{figure}

The sub-array observations resulted in images with size
$128\times 1024$ pixels whereas the full-frame observation resulted in
a 1024$\times 1024$-pixel image; see
Figure~\ref{fig:n49-image}.  The source \rxj\ is very well detected:
in the first sub-array observation a total of 9391 events were
detected in a 3.5-pixel radius and energy range 0.3--10 keV, over the
estimated integration time of 37527 s, while in the second sub-array
observation we detected 11148 counts over 49019 s. In both cases, the
background has not been subtracted.  Background subtraction is tricky
given the high level of background (we will revisit this topic later).
We see that the count rate is  noticeably different between the two visits.

\begin{figure*}
\epsscale{2.0}
\plottwo{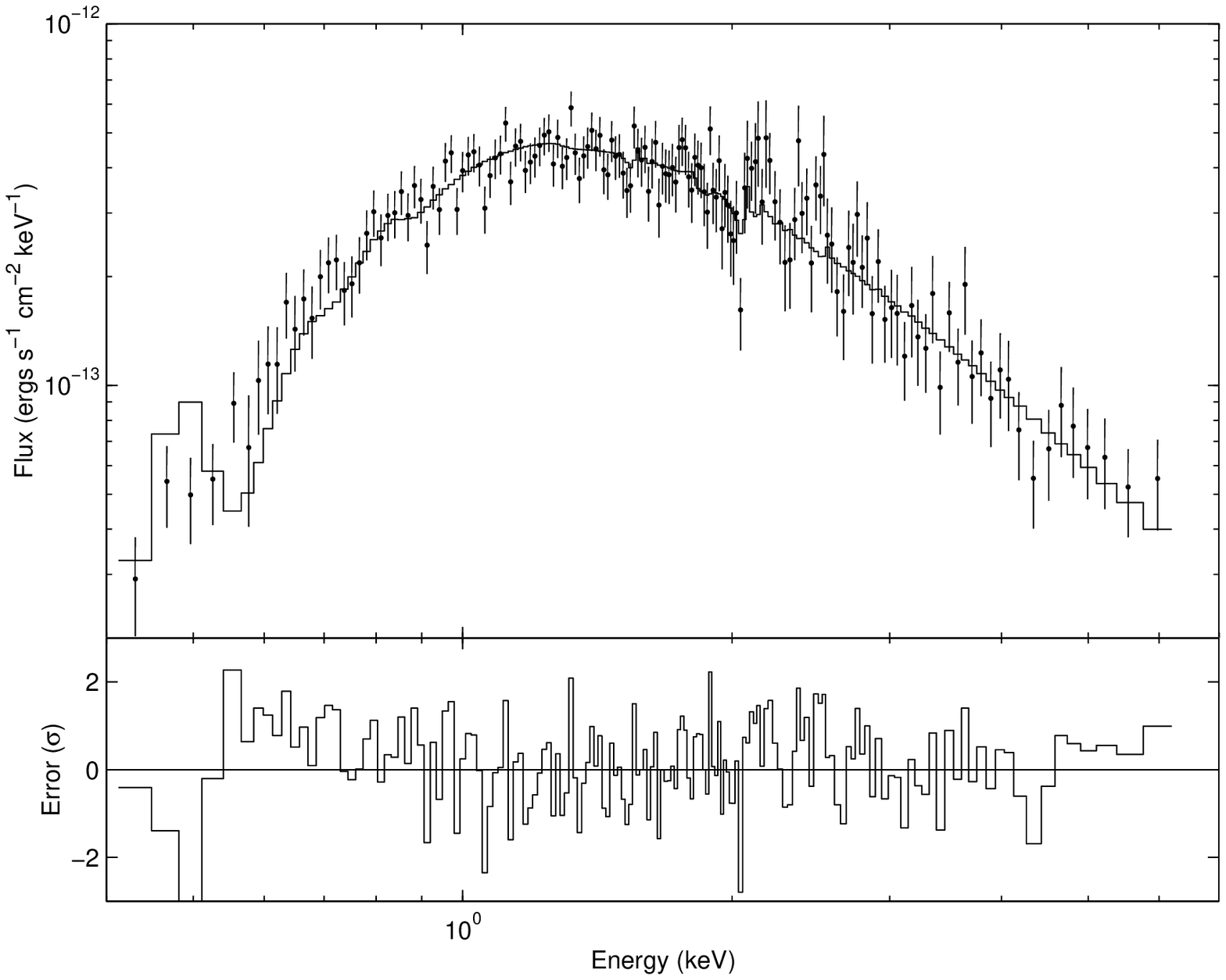}{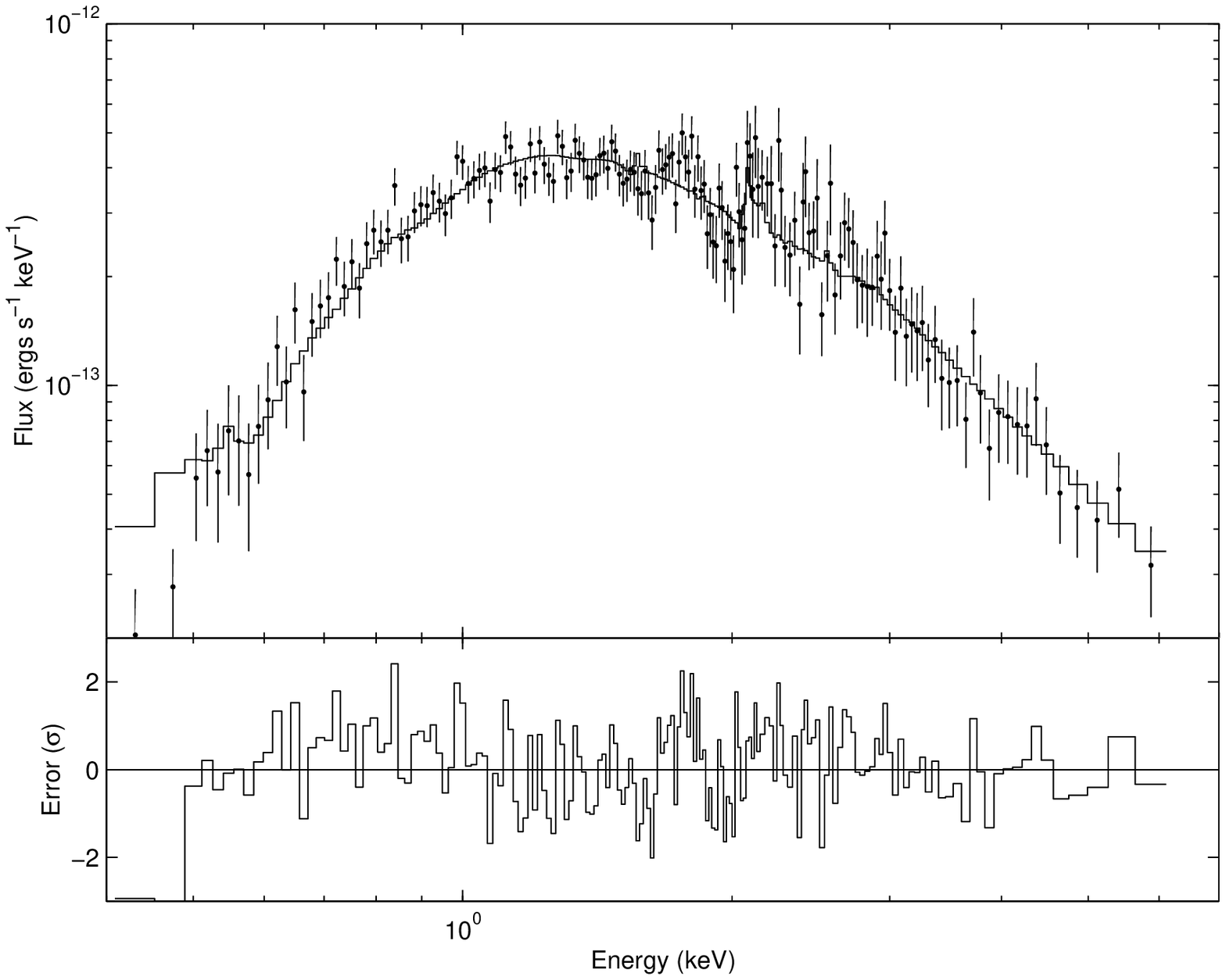}
\caption{\chandra\ spectra of \sgr, from 0.4~keV to 6~keV.  The data
are shown as points, with the best-fit PL+BB (Type b) models (see
Table~\ref{tab:specfit}) shown as the solid lines.  The lower panels
show the residuals, in units of $\sigma$.  The data are ObsId 747
(left) and ObsId 1957 (right).  The fits are in general quite good.
The ``features'' near 1.75~keV and 2.5~keV are from improperly
subtracted nebular emission (from Si and S, respectively), likely due
to the spatial variations of thermal emission from the
supernova remnant N49.
\label{fig:spec}
}
\epsscale{1.0}
\end{figure*}

Having detected the source, we fit a 1-dimensional Gaussian with
$\sigma=0.33$ arcsecond (corresponding to a full-width at half
maximum, FWHM = 0.78 arcsecond) in each axis to the events. The
value of $\sigma$ is comparable to that expected from a point
source and thus \rxj\ is unresolved even at \chandra's exquisite
angular resolution.

After correcting for known aspect
errors\footnote{\url{http://asc.harvard.edu/cal/ASPECT/fix\_offset/fix\_offset.cgi}}
we fit for the position of \sgr\ using an iterative technique.  First
we determined the mean $x$ and $y$ source positions (using
$\sigma$-clipping with a 3-$\sigma$ limit) of the events in a 3-pixel
($1\farcs5$) region around the nominal position of \sgr.  We then used
this new position to refine the center of the source region, and
iterated until the position converged (which occurred in 3--4
iterations depending on the data set).  As can be seen from
Table~\ref{tab:position} the best fit position of \rxj\ is right
ascension $05^{\rm h}26^{\rm m}00.89^{\rm s}$ and declination
$-66^\circ 04^\prime 36\farcs3$ (equinox J2000); the photon
(stochastic) error is negligible and the error is dominated by
0.6-arcsecond systematic error in each coordinate arising from uncertain aspect. This
position can be compared with the \rosat\ position of $05^{\rm
h}26^{\rm m}00.3^{\rm s}$ and $-66^\circ 04^\prime 33\farcs2$ with an
uncertainty of 5-arcsecond (radius).

We inspected the image for evidence of a compact non-thermal nebula ---
a plerion --- but found no evidence for one. However, strong
diffuse emission from N49 is seen. Indeed, at a radius of 2-arcsecond
we clearly detect thermal SNR emission replete with line features: Mg-K
(1.25 keV), Si-K (1.74 keV), S-K (2.31 keV) and Ar-K (2.96 keV).  Such
a spectrum is typical of the emission expected from a middle-aged SNR
(see also Table~\ref{tab:specfit}).

\subsection{Spectral Analysis}

For the spectral analysis, we only used the data from ObsId's 747 and
1957, as these were not affected by photon pileup.  We extracted the
counts from a region around the source position with a 2-pixel
($1\arcsec$) radius for spectral analysis; for the background, we used
an annulus with radius from  2 to 10~pixels (we use an aperture
correction of 8\% to account for the finite extraction aperture, 
determined using \texttt{mkpsf}).
We then used the
\texttt{psextract}\footnote{\url{http://asc.harvard.edu/ciao/threads/psextract/}
} tool to bin the data and generate the necessary response files.  The
spectral data were binned to have 20~counts in each bin.

We fit the data using three models: black-body (BB), power-law (PL),
power-law plus blackbody (PL+BB), all modified by interstellar
absorption \citep[][assuming Solar abundances]{bm92}.  We required that
both observations have the same interstellar absorption column density,
$N_H$.  We tried two types of fits for each model: Type~a is where the
fit parameters were held to be the same over both observations, and
Type~b is where all parameters other than $N_H$ were allowed to
differ.  The results of these four fits (two models and two types) are
shown in Table~\ref{tab:specfit}.

The single blackbody (BB) model produces unacceptable $\chi^2$.  The
fit shows systematic deviations in the following bands: 0.5--0.8 keV,
1.0--1.5 keV and 3.5--7.0 keV. Furthermore, the BB fit results in an
inferred interstellar absorption column, $N_H$, well below that
obtained from analysis of the emission from the supernova remnant (SNR)
N49 (see below).  So we decisively reject the BB model.

As can be seen from Table~\ref{tab:specfit} and Figure~\ref{fig:spec}
the PL and PL+BB models provide acceptable fits.  To determine
statistically which fit is the best, we used an F-test (see
\citet[][p.\ 208]{br92}). This test involves comparison of the
difference in $\chi^{2}$ values (between a given fit and the best fit
model) and the difference in degrees of freedom , to the $\chi^{2}$
value and degrees of freedom for the best-fit (PL+BB, Type b) model.
As seen in Table~\ref{tab:specfit}, complicated models are highly
preferred over the simplest (PL Type a) model: the Type b PL+BB model
is preferred at the 99.97\% confidence level.  This indicates that a
blackbody component is preferred for the fit at the 90\% confidence
level, and that while the power-law indices and blackbody temperatures
are similar across the fits there is a change in absolute flux,
necessitating the Type b model.  This change is likely the result of
the degradation of the ACIS
detectors\footnote{\url{http://asc.harvard.edu/cal/Acis/Cal\_prods/qeDeg/}}.

Separately, we carried out a single temperature MEKA-L model in
\texttt{xspec} of the SNR emission close to \rxj\ and obtained adequate
fit $kT$ of 0.21~keV and $N_H=(6.4\pm 0.1)\times 10^{21}\,$cm$^{-2}$.
(A more detailed analysis of the SNR spectrum is in progress.)

\subsection{Search for Periodicity}
\label{sec:periodicity}
For other SGRs, periodicity has been detected in the quiescent X-ray
emission \citep{kds+98,hlkm+99}. \citet{mrlp96} searched
unsuccessfully for periodicity from \rxj\ in the \rosat\ data, but
their limit of 66\% pulse fraction was not very stringent.

\begin{figure}
\plotone{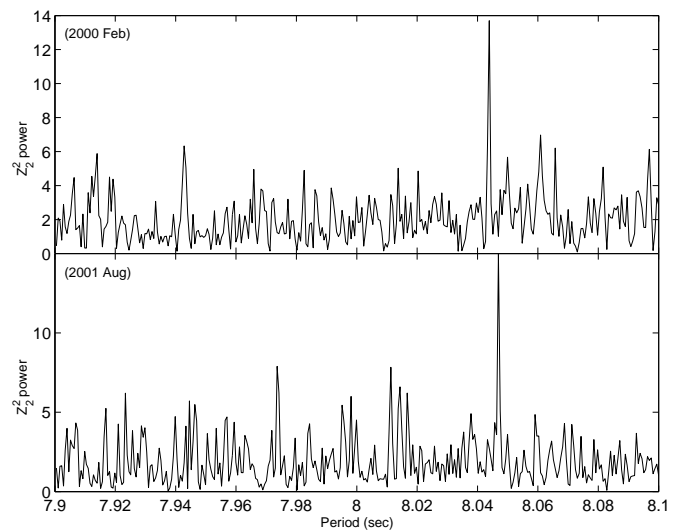}
\caption{$Z_{2}^{2}$ periodograms of ObsID 747 (top) and ObsID 1957
  (bottom) showing the most probable periodicities of 8.0436(2)-s and
  8.0470(2)-s respectively.  The nominal change in period implies a secular
  period derivative $\dot P=\expnt{6.5(5)}{-11}\mbox{ s s}^{-1}$.
}
\label{fig:period}
\end{figure}

We used the well-known statistic $Z_n^2$ \citep*{jrs89} to search for
periodicity.  After transforming the arrival times of the events in
ObsID 747 to the barycenter of the solar system, we added a random
number drawn uniformly from the range [0.0,0.44104] s to remove any
artifacts created by the readout process.  We began searching with the
$Z_{1}^{2}$ statistic around a range encompassing the previously noted
period (7.9--8.1 s) but found no significant peak.  Re-inspecting the
pulsations in the afterglow of 5 March 1979 we noted that the
interpulse gets stronger toward the end of the afterglow of 5 March
1979 \citep{cdpt+80}.  A strong interpulse located 180 degrees in
phase from the main pulse will result in weakening the fundamental and
the second harmonic. Motivated thus we searched with the $Z_{2}^{2}$
statistic which incorporates power from the first harmonic in the
periodogram, and found a peak of moderate significance at
$8.0436(2)$~s (Figure~\ref{fig:period}).

Using this detection as a starting point, we searched for related
periodicities in the data from ObsID 1957.  We find
a peak of similar strength in the $Z_{2}^{2}$ periodogram at
8.0470(2)~s; see Figure~\ref{fig:period}.  Here, though, while the
strength of the peak is similar in the two observations the
significance is higher in the second, as we can restrict the region
searched in period space to those values allowed by the range of
expected period derivatives ($0 \leq \dot P \leq
10^{-10}\mbox{s~s}^{-1}$ or 8.0436--8.0488~s; although we show the full
7.9--8.1~s in Figure~\ref{fig:period} for clarity).  With this
restricted range, the significance of the second periodicity increases
to $\sim 99.98$\%.  The secular spin-down inferred from these two
observations is $\expnt{6.5(5)}{-11}\mbox{ s s}^{-1}$, in the range
found for SGRs and AXPs \citep{hurley99,mereghetti99}.

As can be inferred from the marginal detection, the pulse fraction is
quite low, $F\sim 10\%$ where $F=mean(LC)/min(LC)-1$ where $mean(LC)$
is the mean of the light curve and $min(LC)$ is the minimum of the
light curve.

\section{Discussion}

Here, we report \chandra\ observations of the X-ray counterpart of
\sgr. We have three primary results from these observations: (1) We
have determined an accurate position for \rxj\
(Table~\ref{tab:position}).  (2) We can rule out pure blackbody (BB)
model for the X-ray spectrum.  Instead we find that the best fit model
requires both a BB component and a power-law (PL) component; the photon index,
$\Gamma\sim 3.1$, is steep (Table~\ref{tab:specfit}).  (3) Restricting the period search to a
range of 8-s (and its harmonic) we detect periodicity with $P\sim 8\,$s
in both datasets. If we assume the period evolves secularly then $\dot
P\sim 6.5\times 10^{-11}$ s s$^{-1}$.  We now discuss these points in
more detail.

The accurate position\footnotemark\footnotetext{The position reported
here has been corrected using the latest aspect solutions and has
higher precision than that given in \cite{kkvk+01}.} in conjunction
with \textit{Hubble Space Telescope} (\hst) images enabled us to place
the most stringent limits to the optical emission from \rxj\
\citep{kkvk+01}. These are the best limits to quiescent optical/IR
emission from an SGR. In particular, in \citet{kkvk+01} we
investigated $F_{XR}$, the ratio of the integrated flux in the X-ray
band (i.e.\ $\nu f_\nu$) to that in the optical R band. As noted by
\citet*{hvkk00}, AXPs are distinguished by an unusually large
$F_{XR}\sim 10^4$.  \rxj\ possesses a similarly large $F_{XR}$
\citep{kkvk+01} --- further evidence of commonality between \sgr\ and
the AXPs.

Next, we draw attention to the fact that $\Gamma$ of \rxj\ is decidedly
steeper than the value of $\sim$2 found for the quiescent emission from
other SGRs \citep{hurley99,ktw+01,fkkf01}, but is similar to the
values of 3 to 4 for AXPs \citep{mereghetti99}.  We view this
similarity with considerable interest since AXPs are unique among
neutron stars for their steep spectra.  Furthermore, we note that a
significant fraction of luminosity for both SGRs and AXPs comes out in
the X-ray band. Thus any commonality in the X-ray spectrum takes on
additional importance.  Indeed, spectral dissimilarity is the reason
why the 7.7-s X-ray pulsar 4U 1626$-$67 is not considered to be an AXP
even though this source shares many attributes with AXPs but has a flat
X-ray spectrum \citep{awnk+95}.

The possible detection of periodicity in the quiescent emission with
$P\sim 8$, similar to the value of the period in the afterglow of 5
March 1979 \citep{mgia+79,cdpt+80} is in accord with what has been
seen in other SGRs. In particular, a period of about 5-s was detected
in the afterglow of the giant flare of 27 August 1998 from SGR 1900+14
\citep{fhd+01} and a similar period was also noted in the quiescent
emission \citep{hlkm+99}. 
Returning to \rxj\, if we accept the $\dot P$ (based on only two
epochs) represents the secular period derivative, then the
characteristic age, $P/2\dot P\sim 2,000\,$yr and inferred vacuum
dipole field strength, $B^2=10^{39} P\dot P$, is $B\sim 7\times
10^{14}\,$G. The age is comparable to the estimated age of the SNR
N49, $\sim 5000\,$yr \citep{vblr92} and the inferred $B$ values are
similar to those inferred for other magnetars and AXPs
\citep{kds+98,mereghetti99}.

\section{Ramifications and Speculations}

In the previous section, we summarized our principal observational
results: the broad band spectrum and evidence for periodicity in \rxj, the
X-ray counterpart of \sgr. Here we consider the ramifications of the
broad-band X-ray spectrum, specifically the steep value of the
photon index of the power-law component, $\Gamma\sim 3.1$, intermediate
to the $\Gamma\sim 2 $ of SGRs and $\Gamma \sim 3$ to 4 of AXPs.
There are two interesting
consequences of this finding.

First, the intermediate value of $\Gamma$ is suggestive of
\sgr\ providing an evolutionary link between SGRs and AXPs.  For both
SGRs and AXPs, the PL component has more energy than the BB component;
this is especially true of \rxj\ and AXPs (see below).
This and the continuity in $\Gamma$ lead us to propose that the PL
component is a manifestation of the underlying physical parameter which
determines whether a magnetar is an SGR or an AXP.   Along these lines,
we note that \citet{kgcl+2001} find that the timing noise of AXP
1E~1048.1$-$5937 is considerably worse than those of other AXPs.
Curiously enough, of all the AXPs, this object has the smallest PL
index, $\Gamma \sim 2.5$ and has recently been seen to emit small
bursts \citep{gkw02}.  Thus both \sgr\ and 1E~1048.1$-$5937 appear to
be ``transition'' objects between the two classes.  Furthermore,
\citet{mw2001} find a correlation between spectral hardness
(essentially the PL index) and $\dot P$ (which usually correlates with
timing noise; \citealt{antt94,gk02}).  Thus from these entirely
independent considerations, once again there is a suggestion of
$\Gamma$ being a parameter which varies smoothly from AXPs to SGRs.

Second, the steep value argue that the power law (PL) component
dominates the energy output.  Specifically, as can be seen from
Table~\ref{tab:specfit}, the PL flux, even when restricted to photons
above 0.5 keV, dominates over the black body. We do not know at what
(low) energy the PL component cuts off.  It is clear from the faint
optical flux of \rxj\ \citep{kkvk+01} that the PL component must turn
over somewhere between 0.5 keV (the lowest channel in which we have
some detection) and the optical, and the location of this turn-over
determines the luminosity of \rxj.  For instance, if the PL component
turns over at 50 eV, then the PL flux will be 150 times larger than
the BB flux. The best way (or the only way, to our knowledge) to constrain
the low energy cutoff is by calorimetry via nebular recombination
lines.

Above we have argued that the PL component is a manifestation of the
underlying physical parameter which determines whether a magnetar is
an AXP or SGR. What physical parameter determines $\Gamma$?  One
possibility is the geometry of the
magnetic field. We consider two possibilities. 
AXPs have smooth dipole fields
and SGRs have tangled (multipolar) fields. The latter may then suffer
from frequent magnetic reconnections and thus account for the
super-flares.  The pulse fractions (defined as in
\S\ref{sec:periodicity}) appear to favor this simple idea: AXPs have
large pulse fractions \citep*{opk01}, between 30\% and 70\% (with the
exception of 4U~0142+61 for which the pulse fraction is 10\%) whereas
SGRs have small fractions, 10\% to 20\% for the quiescent counterparts
of \sgr\ (this work), SGR~1900+14 \citep{hlkm+99} and SGR~1806$-$20
\citep{kfkg+02}.  One expects multipole fields to decay more rapidly
compared to dipole fields and thus in this framework, SGRs should be
younger than AXPs. However, the current 
data, namely the association
of three AXPs with SNRs, taken at face value,
seemingly argue for the opposite conclusion. We do recognize that
this inference is based on a small  sample:  six AXPs, three of
have associated SNRs and at most one SNR association for
SGRs (namely, the object of this paper).

Another possibility for differing geometry 
is to invoke large scale twists of a dipole
field with the twist angle being the underlying physical 
parameter \citep*{tlk02}.  In this model, the BB flux arises
both from the heating of the surface due to the decay of strong
magnetar fields \citep{td96,hk98} as well as heating of the surface by
the return current. Resonant cyclotron scattering of these photons by
the magnetosphere is responsible for the PL component. The twist angle could
be the underlying physical parameter that differentiates AXPs from SGRs. 
We refer to \citet{tlk02} for further discussion of this hypothesis.

As noted in \S\ref{sec:introduction} and also above, there are
considerable difficulties in linking AXPs to SGRs via temporal
evolution.  Specifically, the period and period derivatives of AXPs
and SGRs overlap and in are strongly clustered.  Thus, the simplest
interpretation of the overlap of properties is that AXPs and SGRs are
similar objects but in differing ``states''.  As an example, we note
that \sgr\ behaved like a classical SGR from its discovery in 1979
until 1983, but has been silent since then and this may account why
the current spectral properties of \sgr\ are similar to those of AXPs.

We do not know the duty cycle of the two states (AXP and SGR).
If magnetars spend a significant fraction of time in the
AXP state then the radiated energy (assuming say 50 eV low
energy cutoff for the PL component) can be as high as 
$1.2\times 10^{37}\,$ erg s$^{-1}\times 10^4\,{\rm yr} \sim 3\times 10^{48}\,$erg.
The inferred $B$-field value (to supply this energy) is in excess
of $10^{15}\,$G.
As noted in \S\ref{sec:introduction} there is growing evidence for
pulsars with strong $B$-fields, $10^{13} \ale B \ale 10^{14}\,$G
(HBPSRs).  \citet{zh00} have suggested that neutron stars with $B\age
10^{14}\,$G will not exhibit radio pulsations. If so, there may exist
an intermediate group of neutron stars with $10^{14} \ale B \ale
10^{15}$\,G which are neither radio pulsars nor members of the AXP+SGR
family.  Perhaps the nearby X-ray pulsar RBS~1223 \citep{hhss2001} may
be a member of this intermediate group.

\acknowledgements
We have made extensive use of the SIMBAD database and we are grateful
to the astronomers at the Centre de Donn\'ees Astronomiques de
Strasbourg for maintaining this database.  We thank Marten van
Kerkwijk, Chris Thompson, Andrew Melatos and Bing Zhang for helpful
discussions.  DLK thanks the Fannie and John Hertz Foundation for a
fellowship.  Support for this work was provided by the National
Aeronautics and Space Administration (NASA) through {\em Chandra}
award number GO1-2056X issued by the {\em Chandra} X-ray Observatory
Center which is operated by the Smithsonian Astrophysical Observatory
for and on behalf of NASA under contract NAS8-39073.

\bibliographystyle{apj}

\begin{thebibliography}{40}

\bibitem[{Angelini} {et~al.}(1995){Angelini}, {White}, {Nagase}, {Kallman},  {Yoshida}, {Takeshima}, {Becker}, \& {Paerels}]{awnk+95}
{Angelini}, L., {White}, N.~E., {Nagase}, F., {Kallman}, T.~R., {Yoshida}, A.,  {Takeshima}, T., {Becker}, C., \& {Paerels}, F. 1995, \apj, 449, L41

\bibitem[{Arzoumanian} {et~al.}(1994){Arzoumanian}, {Nice}, {Taylor}, \&  {Thorsett}]{antt94}
{Arzoumanian}, Z., {Nice}, D.~J., {Taylor}, J.~H., \& {Thorsett}, S.~E. 1994,  \apj, 422, 671

\bibitem[{Balucinska-Church} \& {McCammon}(1992){Balucinska-Church} \& {McCammon}]{bm92}
{Balucinska-Church}, M. \& {McCammon}, D. 1992, \apj, 400, 699

\bibitem[{Bevington} \& {Robinson}(1992){Bevington} \& {Robinson}]{br92}
{Bevington}, P.~R. \& {Robinson}, D.~K. 1992, {Data reduction and error  analysis for the physical sciences} (New York: McGraw-Hill, 2nd ed.)

\bibitem[{Camilo} {et~al.}(2000){Camilo}, {Kaspi}, {Lyne}, {Manchester},  {Bell}, {D'Amico}, {McKay}, \& {Crawford}]{ckl+00}
{Camilo}, F., {Kaspi}, V.~M., {Lyne}, A.~G., {Manchester}, R.~N., {Bell},  J.~F., {D'Amico}, N., {McKay}, N. P.~F., \& {Crawford}, F. 2000, \apj, 541,  367

\bibitem[{Cline} {et~al.}(1980){Cline}, {Desai}, {Pizzichini}, {Teegarden},  {Evans}, {Klebesadel}, {Laros}, {Hurley}, {Niel}, \& {Vedrenne}]{cdpt+80}
{Cline}, T.~L., {et al.} 1980, \apj, 237, L1

\bibitem[{de Jager} {et~al.}(1989){de Jager}, {Raubenheimer}, \&  {Swanepoel}]{jrs89}
{de Jager}, O.~C., {Raubenheimer}, B.~C., \& {Swanepoel}, J. W.~H. 1989, \aap,  221, 180

\bibitem[{Duncan} \& {Thompson}(1992){Duncan} \& {Thompson}]{dt92}
{Duncan}, R.~C. \& {Thompson}, C. 1992, \apjl, 392, L9

\bibitem[{Evans} {et~al.}(1980){Evans}, {Klebesadel}, {Laros}, {Cline},  {Desai}, {Teegarden}, {Pizzichini}, {Hurley}, {Niel}, \&  {Vedrenne}]{eklc+80}
{Evans}, W.~D., {et al.} 1980, \apj, 237, L7

\bibitem[{Feroci} {et~al.}(2001){Feroci}, {Hurley}, {Duncan}, \&  {Thompson}]{fhd+01}
{Feroci}, M., {Hurley}, K., {Duncan}, R.~C., \& {Thompson}, C. 2001, \apj, 549,  1021

\bibitem[Fox {et~al.}(2001)Fox, Kaplan, Kulkarni, \& Frail]{fkkf01}
Fox, D.~W., Kaplan, D.~L., Kulkarni, S.~R., \& Frail, D.~A. 2001, \apj,  submitted (astro-ph/0107520)

\bibitem[{Gaensler} {et~al.}(2001){Gaensler}, {Slane}, {Gotthelf}, \&  {Vasisht}]{gsgv01}
{Gaensler}, B.~M., {Slane}, P.~O., {Gotthelf}, E.~V., \& {Vasisht}, G. 2001,  \apj, 559, 963

\bibitem[{Gavriil} \& {Kaspi}(2002){Gavriil} \& {Kaspi}]{gk02}
{Gavriil}, F.~P. \& {Kaspi}, V.~M. 2002, \apj, 567, 1067

\bibitem[Gavriil {et~al.}(2002)Gavriil, Kaspi, \& Woods]{gkw02}
Gavriil, P., Kaspi, V.~M., \& Woods, P.~M. 2002, Nature, 419, 142

\bibitem[{Hambaryan} {et~al.}(2002){Hambaryan}, {Hasinger}, {Schwope}, \&  {Schulz}]{hhss2001}
{Hambaryan}, V., {Hasinger}, G., {Schwope}, A.~D., \& {Schulz}, N.~S. 2002,  \aap, 381, 98

\bibitem[{Heyl} \& {Kulkarni}(1998){Heyl} \& {Kulkarni}]{hk98}
{Heyl}, J.~S. \& {Kulkarni}, S.~R. 1998, \apj, 506, L61

\bibitem[{Hulleman} {et~al.}(2000){Hulleman}, {van Kerkwijk}, \&  {Kulkarni}]{hvkk00}
{Hulleman}, F., {van Kerkwijk}, M.~H., \& {Kulkarni}, S.~R. 2000, \nat, 408,  689

\bibitem[Hurley(2000)Hurley]{hurley99}
Hurley, K. 2000, in Gamma-Ray Bursts: 5th Huntsville Symposium, ed. R.~M.  Kippen, R.~S. Mallozi, \& G.~J. Fishman, 763 (astro-ph/9912061)

\bibitem[{Hurley} {et~al.}(1999){Hurley}, {Li}, {Kouveliotou}, {Murakami},  {Ando}, {Strohmayer}, {van Paradijs}, {Vrba}, {Luginbuhl}, {Yoshida}, \&  {Smith}]{hlkm+99}
{Hurley}, K., {et al.} 1999, \apj, 510, L111

\bibitem[{Kaplan} {et~al.}(2002){Kaplan}, {Fox}, {Kulkarni}, {Gotthelf},  {Vasisht}, \& {Frail}]{kfkg+02}
{Kaplan}, D.~L., {Fox}, D.~W., {Kulkarni}, S.~R., {Gotthelf}, E.~V., {Vasisht},  G., \& {Frail}, D.~A. 2002, \apj, 564, 935

\bibitem[{Kaplan} {et~al.}(2001){Kaplan}, {Kulkarni}, {van Kerkwijk},  {Rothschild}, {Lingenfelter}, {Marsden}, {Danner}, \& {Murakami}]{kkvk+01}
{Kaplan}, D.~L., {Kulkarni}, S.~R., {van Kerkwijk}, M.~H., {Rothschild}, R.~E.,  {Lingenfelter}, R.~L., {Marsden}, D., {Danner}, R., \& {Murakami}, T. 2001,  \apj, 556, 399

\bibitem[{Kaspi} \& {Gavriil}(2002){Kaspi} \& {Gavriil}]{kg02}
{Kaspi}, V.~M. \& {Gavriil}, F.~P. 2002, \iaucirc, 7924, 3

\bibitem[{Kaspi} {et~al.}(2001){Kaspi}, {Gavriil}, {Chakrabarty}, {Lackey}, \&  {Muno}]{kgcl+2001}
{Kaspi}, V.~M., {Gavriil}, F.~P., {Chakrabarty}, D., {Lackey}, J.~R., \&  {Muno}, M.~P. 2001, \apj, 558, 253

\bibitem[{Kouveliotou} {et~al.}(1998){Kouveliotou}, {Dieters}, {Strohmayer},  {van Paradijs}, {Fishman}, {Meegan}, {Hurley}, {Kommers}, {Smith}, {Frail},  \& {Murakami}]{kds+98}
{Kouveliotou}, C., {et al.} 1998, \nat, 393, 235

\bibitem[{Kouveliotou} {et~al.}(2001){Kouveliotou}, {Tennant}, {Woods},  {Weisskopf}, {Hurley}, {Fender}, {Garrington}, {Patel}, \& {G{\" o}{\u g}{\"  u}{\c s}}]{ktw+01}
{Kouveliotou}, C., {et al.} 2001, \apjl, 558, L47

\bibitem[{Marsden} {et~al.}(1996){Marsden}, {Rothschild}, {Lingenfelter}, \&  {Puetter}]{mrlp96}
{Marsden}, D., {Rothschild}, R.~E., {Lingenfelter}, R.~E., \& {Puetter}, R.~C.  1996, \apj, 470, 513

\bibitem[{Marsden} \& {White}(2001){Marsden} \& {White}]{mw2001}
{Marsden}, D. \& {White}, N.~E. 2001, \apjl, 551, L155

\bibitem[{Mazets} {et~al.}(1979){Mazets}, {Golentskii}, {Ilinskii}, {Aptekar},  \& {Guryan}]{mgia+79}
{Mazets}, E.~P., {Golentskii}, S.~V., {Ilinskii}, V.~N., {Aptekar}, R.~L., \&  {Guryan}, I.~A. 1979, \nat, 282, 587

\bibitem[Mereghetti(2000)Mereghetti]{mereghetti99}
Mereghetti, S. 2000, in The Neutron Star - Black Hole Connection, ed.  V.~Connaughton, C.~Kouveliotou, J.~{van Paradijs}, \& J.~Ventura (NATO  Advanced Study Institute) (astro-ph/9911252)

\bibitem[{Mereghetti} \& {Stella}(1995){Mereghetti} \& {Stella}]{ms95}
{Mereghetti}, S. \& {Stella}, L. 1995, \apjl, 442, L17

\bibitem[{{\" O}zel} {et~al.}(2001){{\" O}zel}, {Psaltis}, \& {Kaspi}]{opk01}
{{\" O}zel}, F., {Psaltis}, D., \& {Kaspi}, V.~M. 2001, \apj, 563, 255

\bibitem[{Paczynski}(1992){Paczynski}]{paczynski92}
{Paczynski}, B. 1992, Acta Astronomica, 42, 145

\bibitem[{Pivovaroff} {et~al.}(2000){Pivovaroff}, {Kaspi}, \&  {Camilo}]{pkc00}
{Pivovaroff}, M.~J., {Kaspi}, V.~M., \& {Camilo}, F. 2000, \apj, 535, 379

\bibitem[{Rothschild} {et~al.}(1994){Rothschild}, {Kulkarni}, \&  {Lingenfelter}]{rkl94}
{Rothschild}, R.~E., {Kulkarni}, S.~R., \& {Lingenfelter}, R.~E. 1994, \nat,  368, 432

\bibitem[{Thompson} \& {Duncan}(1993){Thompson} \& {Duncan}]{td93}
{Thompson}, C. \& {Duncan}, R.~C. 1993, \apj, 408, 194

\bibitem[{Thompson} \& {Duncan}(1996){Thompson} \& {Duncan}]{td96}
---. 1996, \apj, 473, 322

\bibitem[{Thompson} {et~al.}(2002){Thompson}, {Lyutikov}, \&  {Kulkarni}]{tlk02}
{Thompson}, C., {Lyutikov}, M., \& {Kulkarni}, S.~R. 2002, \apj, 574, 332

\bibitem[{van Paradijs} {et~al.}(1995){van Paradijs}, {Taam}, \& {van den  Heuvel}]{pth95}
{van Paradijs}, J., {Taam}, R.~E., \& {van den Heuvel}, E.~P.~J. 1995, \aap,  299, L41

\bibitem[{Vancura} {et~al.}(1992){Vancura}, {Blair}, {Long}, \&  {Raymond}]{vblr92}
{Vancura}, O., {Blair}, W.~P., {Long}, K.~S., \& {Raymond}, J.~C. 1992, \apj,  394, 158

\bibitem[{Weisskopf} {et~al.}(2000){Weisskopf}, {Tananbaum}, {Van Speybroeck},  \& {O'Dell}]{wtvso2000}
{Weisskopf}, M.~C., {Tananbaum}, H.~D., {Van Speybroeck}, L.~P., \& {O'Dell},  S.~L. 2000, in Proc. SPIE Vol. 4012, p. 2-16, X-Ray Optics, Instruments, and  Missions III, Joachim E. Tr\"{u}mper; Bernd Aschenbach; Eds., Vol. 4012,  2--16

\bibitem[{Zhang} \& {Harding}(2000){Zhang} \& {Harding}]{zh00}
{Zhang}, B. \& {Harding}, A.~K. 2000, \apj, 535, L51

\end{thebibliography}

\begin{deluxetable}{l c c c c}
\tablecaption{Summary of Spectral Fits to \rxj}
\tablewidth{0pt}
\tabletypesize{\small}
\tablehead{
\colhead{Parameter} & \mc{4}{c}{Model Type} \\
 & \mc{2}{c}{PL} & \mc{2}{c}{PL+BB} \\
 & \colhead{Type a} & \colhead{Type b} & \colhead{Type a} & \colhead{Type b}\\
}
\startdata
$N_{H}{\rm (}10^{22}\rm{ cm}^{-2})$  & $0.56(1)$ & $0.56(1)$ &
$0.55(2)$ & $0.54(2)$\\
\tableline
 & \mc{4}{c}{ObsId 747} \\
$\Gamma$ & $3.06(3)$ & $3.06(4)$ & $3.14(8)$ & 3.1(1) \\
$\mbox{PL norm}$ $(10^{-3} \mbox{ s}^{-1}\mbox{ cm}^{-2}\mbox{
keV}^{-1})$ & $1.18(5)$ & $1.22(5)$ & $1.08(8)$ & 1.13(8)\\
PL $f_{X}$\tablenotemark{d} $(10^{-12}\mbox{ ergs s}^{-1}\mbox{ cm}^{-2})$ & 1.17 &
1.23 & 1.01 & 1.12\\
PL $f_{X}^{u}$ $(10^{-12}\mbox{ ergs s}^{-1}\mbox{ cm}^{-2})$ & 3.56 &
3.68 & 3.26 & 3.39\\
$kT^{\infty}$ (keV) & \nodata & \nodata & $0.53(6)$ & 0.6(1) \\
$R^{\infty}_{\rm BB}$ (km)\tablenotemark{b} & \nodata & \nodata &
$2.6(5)d_{50}$ & $2(1)d_{50}$\\
BB $f_{X}$ $(10^{-12}\mbox{ ergs s}^{-1}\mbox{ cm}^{-2})$ & \nodata &
\nodata & 0.14 & 0.10 \\
BB $f_{X}^{u}$ $(10^{-12}\mbox{ ergs s}^{-1}\mbox{ cm}^{-2})$ & \nodata &
\nodata & 0.22 & 0.13\\
$f_{X}^{u}$ $(10^{-12}\mbox{ ergs s}^{-1}\mbox{ cm}^{-2})$ & 3.56 &
3.68 & 3.42 & 3.52\\
$L_{X}$ $(10^{36}\mbox{ ergs s}^{-1})$\tablenotemark{b} &
$1.01d_{50}^{2}$\tablenotemark{a} &
$1.04d_{50}^{2}$ & $0.97d_{50}^{2}$\tablenotemark{a} &
$1.00d_{50}^{2}$\\[0.1in]
\tableline
 & \mc{4}{c}{ObsId 1957} \\
$\Gamma$ & $3.06(3)$\tablenotemark{a} & $3.06(4)$ &
$3.14(8)$\tablenotemark{a} & 3.12(8)\\
$\mbox{PL norm}$ $(10^{-3} \mbox{ s}^{-1}\mbox{ cm}^{-2}\mbox{
keV}^{-1})$& $1.18(5)$\tablenotemark{a} & $1.14(4)$&
$1.08(8)$\tablenotemark{a} & $0.98(8)$ \\
PL $f_{X}$ $(10^{-12}\mbox{ ergs s}^{-1}\mbox{ cm}^{-2})$ &
1.17\tablenotemark{a} &
 1.14 &
1.01\tablenotemark{a} & 0.95\\
PL $f_{X}^{u}$ $(10^{-12}\mbox{ ergs s}^{-1}\mbox{ cm}^{-2})$ &
3.57\tablenotemark{
a} & 3.46
& 3.26\tablenotemark{a} & 2.95\\
$kT^{\infty}$ (keV) & \nodata & \nodata &
$0.53(6)$\tablenotemark{a} & 0.48(5)\\
$R^{\infty}_{\rm BB}$ (km)\tablenotemark{b} & \nodata & \nodata &
$2.6(5)d_{50}$\tablenotemark{a} & $3(1)d_{50}$ \\
BB $f_{X}$ $(10^{-12}\mbox{ ergs s}^{-1}\mbox{ cm}^{-2})$ & \nodata &
\nodata & 0.14\tablenotemark{a} & 0.16\\
BB $f_{X}^{u}$ $(10^{-12}\mbox{ ergs s}^{-1}\mbox{ cm}^{-2})$ & \nodata &
\nodata & 0.22\tablenotemark{a} & 0.28\\
$f_{X}^{u}$ $(10^{-12}\mbox{ ergs s}^{-1}\mbox{ cm}^{-2})$
&3.57\tablenotemark{a}&
3.46& 3.42\tablenotemark{a}
& 3.26 \\
$L_{X}$ $(10^{36}\mbox{ ergs s}^{-1})$\tablenotemark{b} &
$1.01d_{50}^{2}$\tablenotemark{a} &
$0.98d_{50}^{2}$ & $0.87d_{50}^{2}$\tablenotemark{a} &
$0.92d_{50}^{2}$\\[0.1in]
\tableline
DOF & 332 & 330 & 330 & 326\\
$\chi^{2}/{\rm DOF}$ & 1.06 & 1.03 & 1.03 & 1.00\\
P(PL$_{a}$)\tablenotemark{c} & \nodata & \expnt{3}{-3} & \expnt{3}{-3} &
\expnt{3}{
-4} \\
\enddata
\tablecomments{All fluxes and luminosities are in the 0.5--10~keV
  range.  Values in parentheses 
  are 1-$\sigma$ statistical uncertainties.}
\tablenotetext{a}{Fixed to be the same as the corresponding value for
ObsId 747.}
\tablenotetext{b}{At a distance of $50 d_{50}$~kpc.}
\tablenotetext{c}{Probability that the Type a PL model is preferred
over the specified model.}
\tablenotetext{d}{The power law (PL) normalization is at 1.0~keV.}
\label{tab:specfit}
\end{deluxetable}

\end{document}